\documentclass[preprint]{aastex}

\usepackage{subfigure}
\usepackage{float}
\usepackage{color}

\slugcomment{Version: 2008 Oct 29}

\def\Gzero   {G35.20$-$0.74}
\def\Gone    {G35.20$-$1.74}
\def\Jfivex  {J1855+0215}
\def\Jfive   {J1855+0251}
\def\Jthree  {J1903+0145}
\def\Jfour   {J1904+0110}
\def\Jseven  {J1907+0127}

\def\VLSR    {$V_{\rm LSR}$}
\def\deg     {$^\circ$}
\def\kms     {km~s$^{-1}$}
\def\masy    {mas~y$^{-1}$}
\def\mjybeam {mJy~beam$^{-1}$}
\def\etal    {et al.}

\begin{document}

\title{Trigonometric Parallaxes of Massive Star Forming Regions: IV. \\
       \Gzero\ and \Gone}

\author{B. Zhang\altaffilmark{1,5}, X. W. Zheng\altaffilmark{1},
 M. J. Reid\altaffilmark{2}, K. M. Menten\altaffilmark{3},
 Y. Xu\altaffilmark{3,6}, L. Moscadelli\altaffilmark{4},
 A. Brunthaler\altaffilmark{3}}

\altaffiltext{1}{Department of Astronomy, Nanjing University, Nanjing
    210093, China}

\altaffiltext{2}{Harvard-Smithsonian Center for Astrophysics, 60
 Garden Street, Cambridge, MA 02138, USA}

\altaffiltext{3}{Max-Plank-Institut f\"ur Radioastronomie, Auf dem H\"ugel
69, 53121 Bonn, Germany}

\altaffiltext{4}{INAF, Osservatorio Astrofisico di Arcetri, Largo E. Fermi 5, 50125 Firenze, Italy}

\altaffiltext{5}{Shanghai Astronomical Observatory, Chinese Academy of
Sciences, Shanghai 200030, China}

\altaffiltext{6}{Purple Mountain Observatory, Chinese Academy of
Sciences, Nanjing 210008, China}

\begin{abstract}

We report trigonometric parallaxes for the high-mass star forming
regions \Gzero\ and \Gone, corresponding to distances of
$2.19^{+0.24}_{-0.20}$~kpc and $3.27^{+0.56}_{-0.42}$~kpc,
respectively. The distances to both sources are close to their near
kinematic distances and place them in the Carina-Sagittarius spiral
arm. Combining the distances and proper motions with observed radial
velocities gives the locations and full space motions of the star
forming regions.  Assuming a standard model of the Galaxy, \Gzero\ and
\Gone\ have peculiar motions of $\approx 13$~\kms\ and $\approx
16$~\kms\ counter to Galactic rotation and $\approx9$~\kms\ toward the
North Galactic Pole.

\end{abstract}

\keywords{ techniques: interferometric --- astrometry --- spiral arm:
  Sagittarius --- distances --- individual (\Gzero, \Gone)}

\section{Introduction}

\Gzero\ (IRAS 18556+0136) and \Gone\ (IRAS 18592+0108) are high-mass
star forming regions.  They both are home to strong methanol (CH$_3$OH)
masers at 12 GHz and were selected as targets for our large project
with the NRAO \footnote{The National Radio Astronomy Observatory is a
facility of the National Science Foundation operated under cooperative
agreement by Associated Universities, Inc } Very Long Baseline Array
(VLBA) to study the spiral structure and kinematics of the Milky Way.
Details of this project are discussed in \citet{Reid08a}, hereafter
called Paper I.

The kinematic distance of \Gzero\ is about 3.3 kpc, based a CO line
velocity of 35~\kms\ \citep{solomon87}.  This star forming region is
home to a massive protostar with a jet driven outflow, rarely observed
towards massive young stellar objects. Sub-arcsecond resolution VLA
observations of the region by \citet{gibb03} at 3.5 and 6.0 cm show
that three concentrations of radio continuum emission break up into 11
individual sources all lying along the outflow.  As for most high-mass
star forming regions, \Gzero\ contains water and OH masers
\citep{forster89}, methanol masers \citep{caswell95}, and many thermal
lines from molecules like CH$_{3}$CN \citep{kalenskii00}, HCO+ and HCN
\citep{gibb03} have been detected.  This region was also observed
recently at mid-infrared wavelengths by \citet{buizer06}. These
observations reveal an extended source with a cometary shape with the
masers concentrated around the head.

The \Gone\ star forming region also contains water, OH and methanol
masers, and a large cometary UCHII region.  The UCHII region, known as
W48A, lies about 20\arcsec\ from the water and OH masers.
Mid-infrared images \citep{buizer04} show a bright point source at the
location of the water masers as well as the UCHII region to the
south-east.  Kinematic distances to \Gone, from various molecular
lines associated with W48A
\citep{Genzel77,Palagi93,Churchwell90,Braz83,Vallee90}, range between
3.0 and 3.4 kpc.

Starting at the Sun, a ray toward Galactic longitude 35\deg\ first
crosses the Carina--Sagittarius spiral arm, then the Crux--Scutum arm,
before recrossing the Carina--Sagittarius arm and then crossing the
Perseus arm. Thus distances are crucial to establish to which arm
\Gzero\ and \Gone\ belong.  Given the large kinematic anomalies
measured for some massive star forming regions, for example for W3OH
\citep{Xu06a, Hachi06}, it is important to measure distances to these
and other sources by direct methods such as trigonometric parallax.

\section{Observations and Calibration Procedures}

On 2005 October 30, 2006 April 7 and October 7, and 2007 April 16, we
observed methanol masers at 12 GHz toward \Gzero\ and \Gone\ with
8-hour tracks on the VLBA. Since, for these sources, the declination
parallax signature is considerably smaller than for right ascension, we
scheduled the observations so as to maximize the right ascension
parallax offsets.  We observed several extragalactic radio sources as
background references which provide independent checks for parallax
solutions.  Table~1 lists the positions, intensities, source
separations, observed radial velocities and beam sizes.

\begin{table}[H]
  \caption[]{Positions and Brightnesses \label{tab:src_pos}}
  \begin{center}
  \begin{tabular}{cccccccc}
\hline \hline
Source      &  R.A. (J2000)  & Dec. (J2000)   & $T_b$    &  $\theta_{sep}$   & P.A. & \VLSR & Beam \\
            & (h~~~m~~~s)  & (\degr~~~\arcmin~~~\arcsec)  & (Jy/b)   &  (\degr) & (\degr)  & (\kms) & (mas) \\
\hline
\Gzero ...  &  18 58 13.0517 & +01 40 35.674   &  0.7 $-$ 0.9     &  ...  & ...  & 31   & 1.2 \\
\Jfivex ... &  18 55 00.1130 & +02 15 41.100   &   0.03           &  1.0  & $-$54& ...  &... \\
\Jfive ...  &  18 55 35.4365 & +02 51 19.562   &   0.20           &  1.4  & $-$29& ...  & ...\\
\Jseven ... &  19 07 11.9963 & +01 27 08.963   &   0.10           &  2.3  & +96  & ...  & ...\\
            &                &                 &                  &       &      &      & \\
\Gone ...   &  19 01 45.5364 & +01 13 32.545   &  3.5 $-$ 9.5     &  ...  & ...  & 43   & 1.2 \\
\Jthree ... &  19 03 53.0632 & +01 45 26.306   &   0.18           &  0.8  & +45  & ...  & ...\\
\Jfour ...  &  19 04 26.3978 & +01 10 36.696   &   0.33           &  0.7  & +94  & ...  & ...\\
\Jseven ... &  19 07 11.9963 & +01 27 08.963   &   0.30           &  1.4  & +80  & ...  & ...\\
\Jfive ...  &  18 55 35.4365 & +02 51 19.562   &   0.50           &  2.2  & $-$43& ...  & ... \\
\hline
  \end{tabular}
  \end{center}
\small{Note. --- The fourth and fifth columns give the peak
brightnesses ($T_b$) at 12 GHz and their separations ($\theta_{sep})$
and position angles (P.A.) east of north between maser and reference
sources. The sixth column gives the Local Standard of Rest (LSR)
velocities of the masers. The last column gives the FWHM size of the
Gaussian restoring beam. Calibrators \Jfive~\citep{Fomalont03} and
\Jseven~\citep{Petrov05} are from the VLBA Calibrator Survey, the other
calibrators are from \citet{Xu06b}. }
\end{table}

Our general observing setups and calibration procedures are described
in Paper I, and here we discuss only aspects of the observations that
are specific to \Gzero\ and \Gone.  We observed two ICRF sources,
J1800+3848 and J1800+7828 \citep{Ma98}, near the beginning, middle and
end of the observations in order to monitor delay and electronic phase
differences among the observing bands.  The right and left-circularly
polarized emission from the 12 GHz methanol masers toward \Gzero\ and
\Gone\ were observed in 4 MHz bands centered at LSR velocities of
31~\kms\ and 43~\kms, respectively.

\begin{figure}[H]
  \centering \includegraphics[angle=-90,scale=0.60]{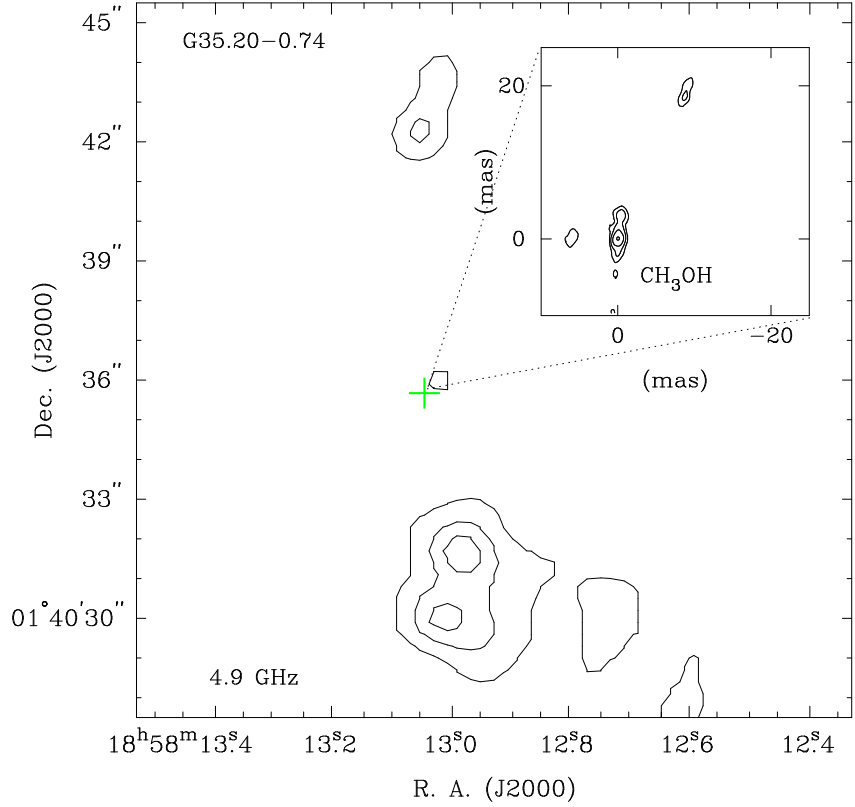}
  \caption{Images of 4.9 GHz radio continuum emission
associated with the star-forming region \Gzero, generated from archival
VLA data (program AH241), and the velocity-integrated maser emission
(inset). The position of the methanol masers is designated with a plus
($+$) sign. Contour levels are linearly spaced at 0.45 mJy for the
continuum emission; they start at 0.06 Jy beam$^{-1}$ \kms\ and
increase by factors of 2 for the maser emission.
  \label{fig:g350_con}}
\end{figure}

\begin{figure}[H]
  \centering \includegraphics[angle=-90,scale=0.60]{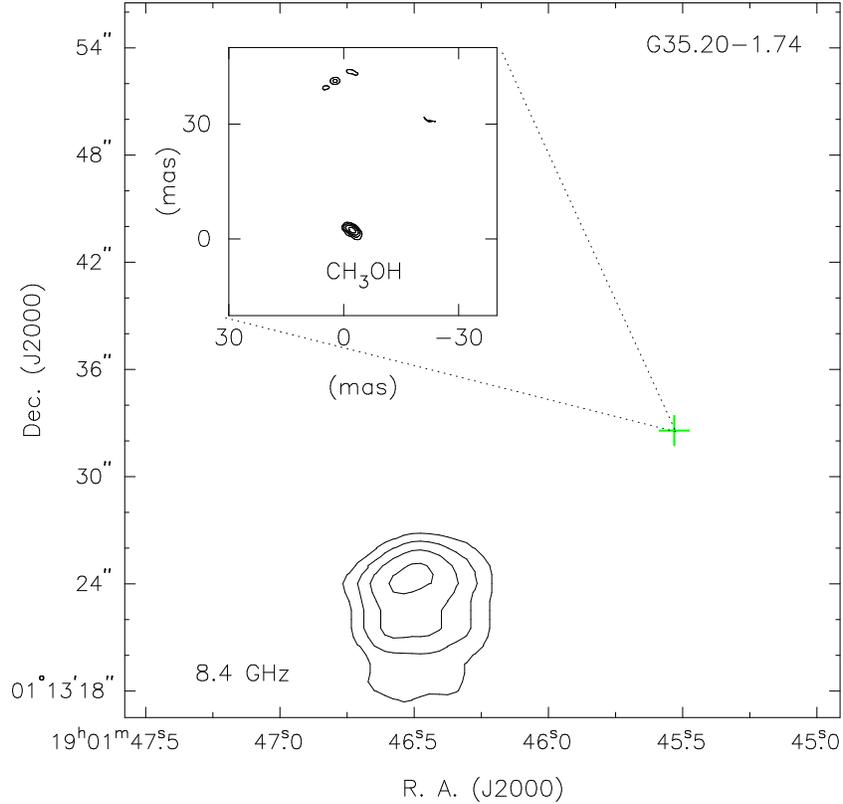}
  \caption{Images of 8.4 GHz radio continuum emission
associated with the star-forming region \Gone, generated from archival
VLA data (program AK477), and the velocity-integrated maser emission
(inset). The position of the methanol masers are shown as a plus ($+$)
sign. Contour levels start at 66 \mjybeam\  and increase by factors of
2 for the continuum emission; they start at 0.27 Jy beam$^{-1}$ \kms\
and increase by factors of 2 for the maser emission.
  \label{fig:g351_con}}
\end{figure}

The strongest maser spots at \VLSR\ of 28.3~\kms\ and 41.5~\kms\ served
as the phase-references for \Gzero\ and \Gone, respectively.  The
point-source response functions (dirty beams) had FWHMs of 1.9 by
0.8~mas at a PA of $-8$\deg\ east of north. After experimenting with
different restoring beams, we adopted a circular restoring beam with a
1.2 mas FWHM for both maser sources and background sources.
Figures~\ref{fig:g350_con} and \ref{fig:g351_con} show radio continuum
(from VLA archive data) and velocity-integrated methanol maser emission
from these star-forming regions.

\section{Parallax and Proper Motion Results}

We fitted elliptical Gaussian brightness distributions to strong maser
spots and the extragalactic radio sources for all four epochs.  The
change in position of each maser spot relative to each background
radio source was modeled by the parallax sinusoid in both coordinates,
determined by one parameter (the parallax), and a secular proper
motion in each coordinate.

\subsection{\label{ses:g350}\Gzero}

\begin{figure}[H]
  \centering
  \includegraphics[angle=-90,scale=0.70]{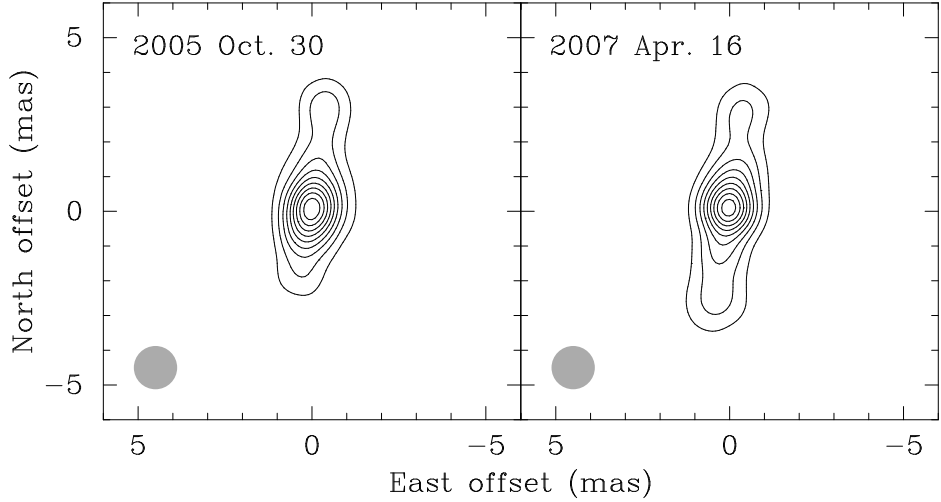}
  \caption{Images of reference maser source at $V_{LSR}$ = 28.3 km
s$^{-1}$ in \Gzero\ at the first and last epoch.  Observation dates are
indicated in the upper left corner of each panel.  The restoring beam
is in the lower left corner. Contour levels are spaced linearly at 90
\mjybeam\ .
  \label{fig:g350_maser}}
\end{figure}

\begin{figure}[H]
  \centering
  \includegraphics[angle=-90,scale=0.70]{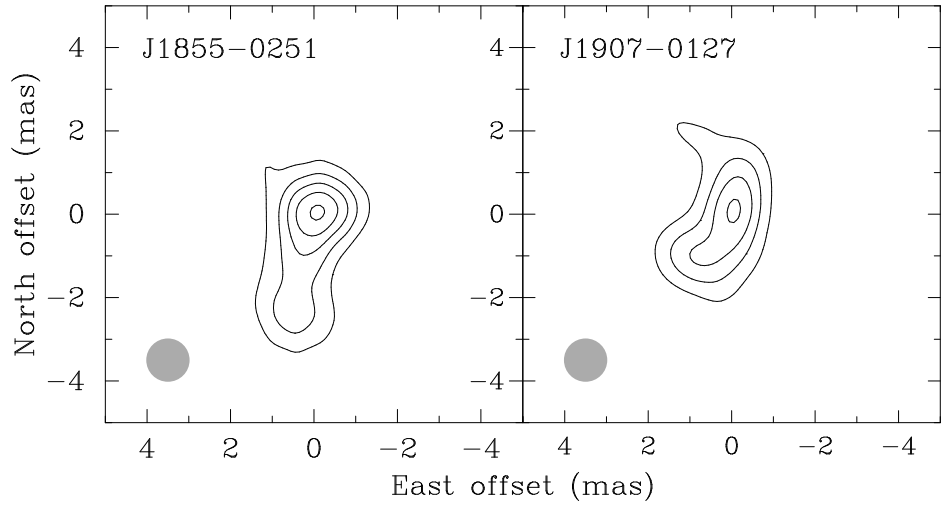}
  \caption{Images of extragalactic radio sources used for the
parallax measurements of \Gzero\ from the first epoch observation
on 2005 Oct 30. Source names are in the upper left corner of each
panel. Contour levels are spaced linearly at 7.0 \mjybeam\  for
\Jfive\ and 10.0 \mjybeam\  for \Jseven.
  \label{fig:g350_qso}}
\end{figure}

Fig.~\ref{fig:g350_maser} shows maser reference channel images for
\Gzero\ from the first and last epochs.  Imaging sources near zero
declination is generally problematic for radio interferometers, and we
suspect that the low-level symmetric structures seen in the \Gzero\
images are caused by small calibration errors. However, since the
parallax measurement comes almost exclusively from the East-West data,
this should not be significant problem.  Keeping this in mind, one can
see that the emission appears dominated by a single compact component,
and there is no significant variation over the 1.5 year time span of
our observations.  One of the three background sources (\Jfivex) proved
to be very weak, and we used only \Jfive\ and \Jseven\ in the parallax
fits. Fig.~\ref{fig:g350_qso} provides the images of background radio
sources used for the fitting of parallax and proper motion for \Gzero.

\begin{figure}[H]
  \centering
    \includegraphics[angle=-90,scale=0.65]{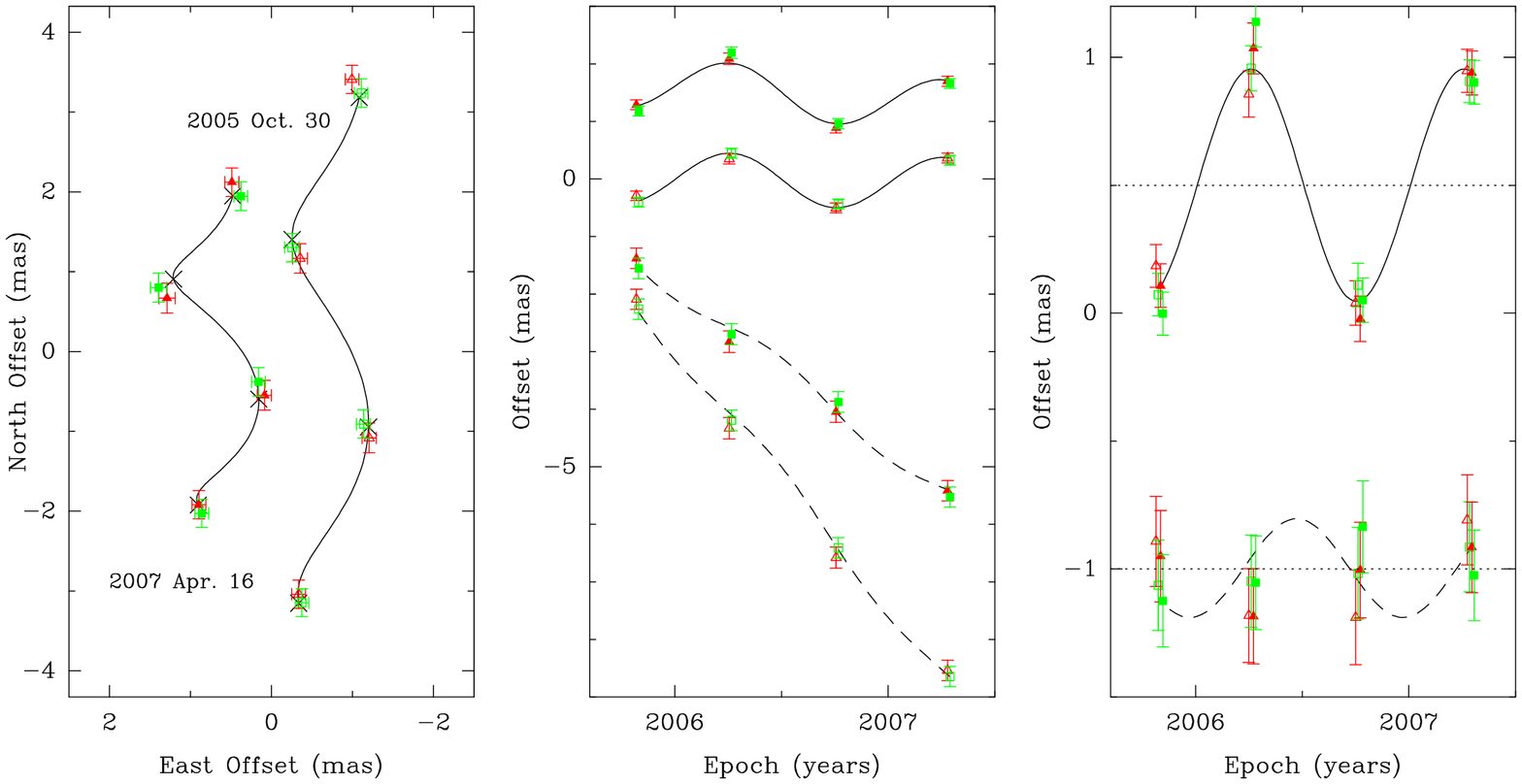}
    \caption{Parallax and proper motion data and best-fitting models
for \Gzero. {\it Filled } and {\it opened} markers are for maser spots
at \VLSR\ of 27.9~\kms\ and 27.5~\kms, respectively. Plotted are
positions of both maser spots relative to the two extragalactic radio
sources: \Jfive\ ({\it triangles}) and \Jseven\ ({\it squares}).  {\it
Left Panel}: Positions with superposed curves representing the modeled
maser track on the sky with the first and last epochs labeled. The
expected positions from the parallax and proper motion fit are
indicated ({\it crosses}). Middle Panel: Eastward ({\it solid lines})
and northward ({\it dashed lines}) offsets and best-fitting models
versus time. Data for the eastward and northward positions are offset
vertically and small time shifts have been added to the data for
clarity.  {\it Right Panel}: Same as the {\it middle panel}, except the
best-fitting proper motions have been removed, allowing all data to be
overlaid and the effects of only the parallax seen. The {\it solid
(dashed) line} shows the combined fitted eastward (northward) parallax
curve.
  \label{fig:g350_para}}
\end{figure}

\begin{table}[H]
  \caption{\Gzero\ Parallax \& Proper Motion Fits}
  \label{tab:g350_para}
  \begin{center}
  \begin{tabular}{ccccc}
\hline\hline
Maser \VLSR & Background  & Parallax         &  $\mu_{x}$  &  $\mu_{y}$ \\
(\kms)      & Source   &  (mas)              &  (\masy)      &  (\masy)     \\
\hline
 27.9 ...   & \Jfive   & 0.411 $\pm$  0.057  &  $-0.07 \pm 0.10$  &  $-4.50 \pm 0.33$ \\
 27.9 ...   & \Jseven  & 0.426 $\pm$  0.037  &  $-0.05 \pm 0.06$  &  $-4.47 \pm 0.10$ \\

 27.5 ...   & \Jfive   & 0.506 $\pm$  0.083  &  $-0.37 \pm 0.14$  &  $-2.83 \pm 0.27$ \\
 27.5 ...   & \Jseven  & 0.522 $\pm$  0.094  &  $-0.35 \pm 0.16$  &  $-2.80 \pm 0.25$ \\
            &          &                     &                    &                   \\
 27.9 ...   & combined & 0.456 $\pm$  0.045  &  $-0.07 \pm 0.08$  &  $-4.47 \pm 0.16$ \\
 27.5 ...   &          &                     &  $-0.29 \pm 0.08$  &  $-2.79 \pm 0.16$ \\
\hline
   \end{tabular}
   \end{center}
\small{Note. --- Combined fit used a single parallax parameter for both
maser spots relative to the background sources, but a different proper
motion was allowed for the two maser spots to allow for internal maser
motions.  Uncertainties are formal errors adjusted to give $\chi^2$ per
degree of freedom of unity, except for the ``combined'' parallax
uncertainty which has been increased to allow for the difference
between the parallaxes of the two maser spots.}
\end{table}

In order to measure the parallax and proper motion of \Gzero, we fitted
the two brightest maser spots, at \VLSR\ = 27.9 and 27.5~\kms, and the
two background radio sources, \Jfive\ and \Jseven, for all four epochs.
In Fig.~\ref{fig:g350_para}, we plot the positions of these two maser
spots relative to the two background radio sources, with superposed
curves representing the model maser tracks across the sky. The fitting
results, listed in Table~\ref{tab:g350_para}, reveal that the
individually measured parallaxes and proper motions for two maser spots
in \Gzero\ are somewhat different.  However, the results using the two
background sources are reasonably consistent. This suggests that time
variable maser structure in one or both spots, and not atmospheric
calibration, may limit the parallax accuracy. Thus, we have increased
the parallax uncertainty for the ``combined'' solution to half the
average difference between the fits for the two maser spots.  The
combined parallax of the two astrometric maser spots is $0.456 \pm
0.045$~mas, corresponding to a distance of $2.19^{+0.24}_{-0.20}$~kpc.

The average proper motion of the two maser spots is $-0.18\pm 0.06$
\masy\ toward the East and $-3.63 \pm 0.11$ \masy\ toward the North.
The difference in the northward proper motions of the two maser spots
of $\pm1.7$ \masy\ ($\pm18$ \kms) is unusually large compared to other
sources given in Papers I -- V \citep{Reid08a, Moscadelli08, Xu08,
Zhang08, Brunthaler08}. While such an internal motion is not
impossible, we suspect that the difference may be a result of the poor
north-south beam of the interferometer for this near zero Declination
source.  At the distance implied by the parallax measurement, the
average proper motions correspond to $-1.9$ \kms\ and $-37.7$ \kms\
eastward and northward, respectively. Completing the 3-D space
velocity, the average LSR velocity of the spots is 27.7 \kms, which
corresponds to a heliocentric radial velocity of 10.7 \kms.

\subsection{\Gone}

\begin{figure}[H]
  \centering
    \includegraphics[angle=-90,scale=0.70]{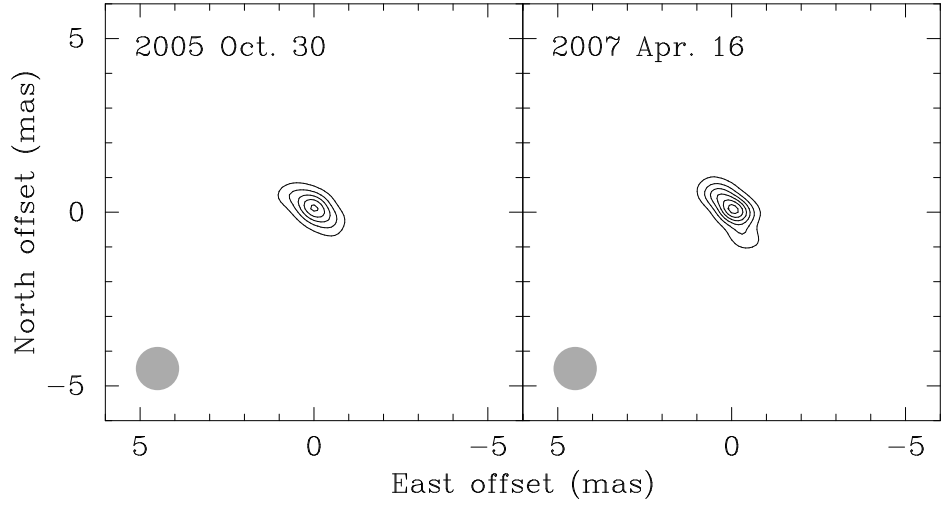}
    \caption{Images of reference maser source at \VLSR\ = 41.5 \kms\
in \Gone\ at the first and last epochs.  The epochs are indicated in
the upper left corner of each panel.  The restoring beam is in the
lower left corner. Contour levels are spaced linearly at 1.2 and 1.5
Jy/beam, respectively.
   \label{fig:g351_maser}}
\end{figure}

\begin{figure}[H]
  \centering
    \includegraphics[angle=-90,scale=0.70]{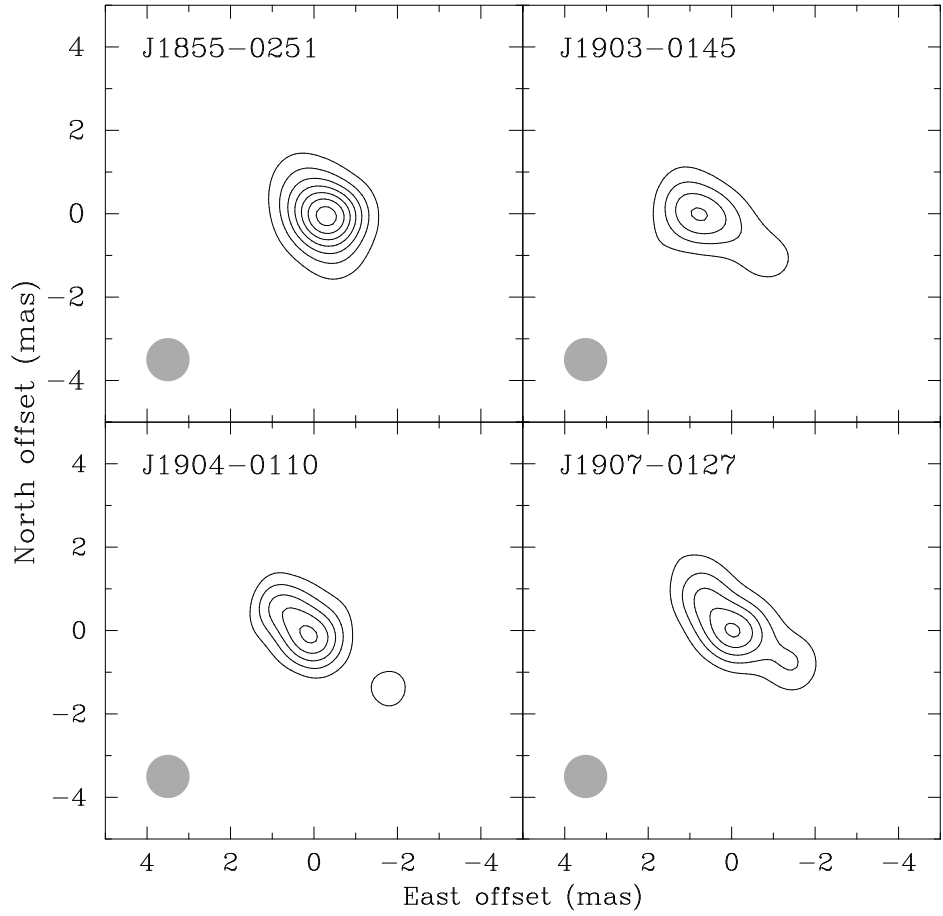}
    \caption{Images of background continuum sources used for the
parallax measurements of \Gone\ at the first epoch.  Source names are
in the upper left corner and restoring beams are in the lower left
corner of each panel.  Contour levels are spaced linearly at increments
of 8.5 \mjybeam\ for \Jfive, 9.5 \mjybeam\  for \Jthree, and 15
\mjybeam\ for \Jfour\ and \Jseven.
    \label{fig:g351_qso}}
\end{figure}

\begin{figure}[H]
  \centering
    \includegraphics[angle=0,scale=0.85]{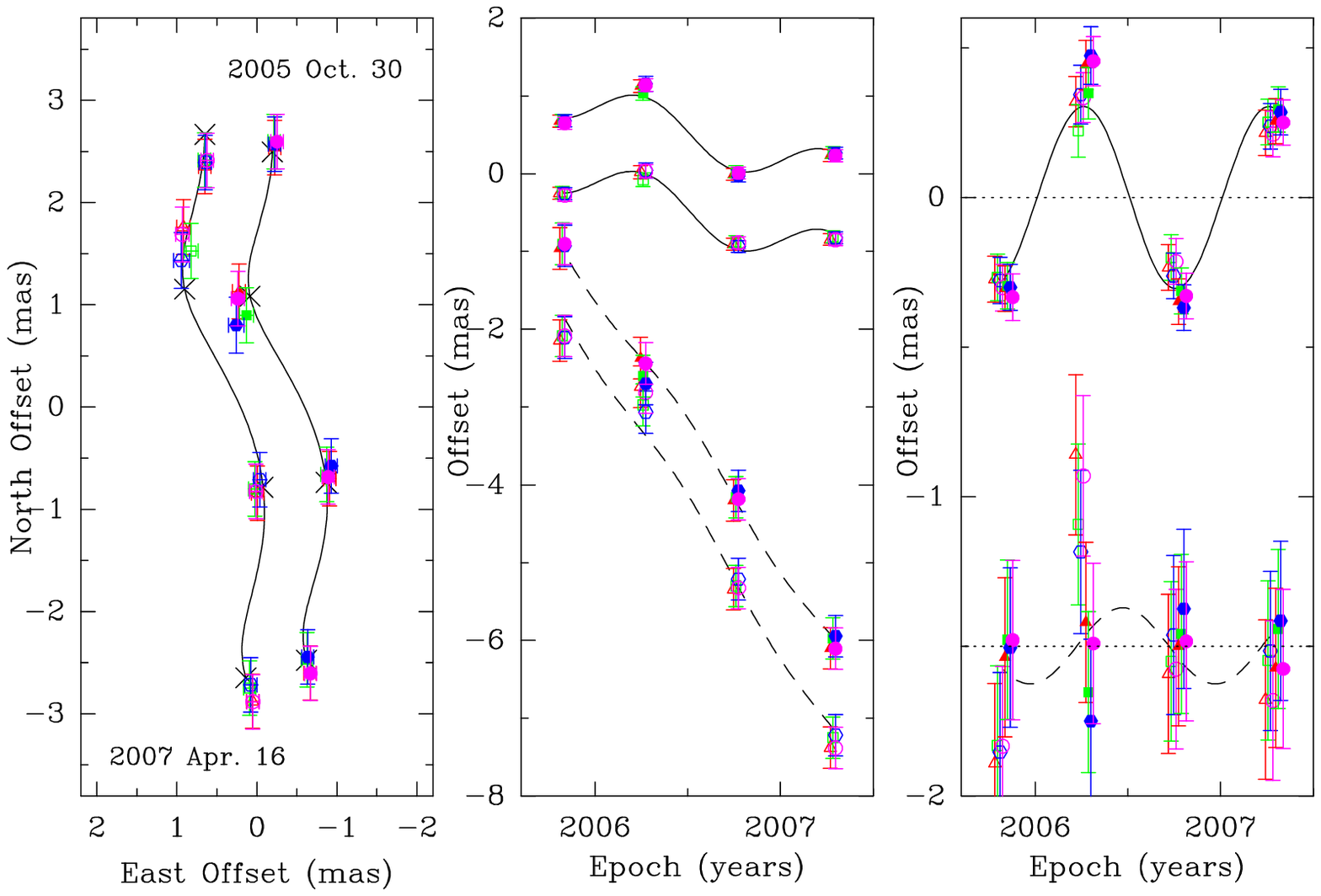}
    \caption{Parallax and proper motion data and fits for \Gone.  {\it
Filled} and {\it opened} markers are for maser spot at \VLSR\ of
41.9~\kms\ and 41.5~\kms, respectively.  Plotted are position
measurements of both maser spots relative to the four background
sources: \Jfive\ ({\it circles}), \Jthree\ ({\it triangles}), \Jfour\
({\it squares}), \Jseven\ ({\it hexagons}).  {\it Left Panel}:
Positions on the sky with the first and last epochs labeled. Data for
the maser spot are offset horizontally for clarity. The expected
positions from the parallax and proper motion fit are indicated ({\it
crosses}). Middle Panel: Eastward ({\it solid lines}) and northward
({\it dashed lines}) positions and best fitting models versus time.
Data for the eastward and northward positions are offset vertically and
small time shifts have been added to the data for clarity. {\it Right
Panel}: Same as the {\it middle panel}, except the best fit proper
motions have been removed, allowing all data to be overlaid and the
effects of only the parallax seen. The {\it solid (dashed) line} shows
the combined fitted eastward (northward) parallax curve.
  \label{fig:g351_para}}
\end{figure}

Figure~\ref{fig:g351_maser} displays images of the maser reference
channel for \Gone\ from the first and last epoch.  The emission
appears dominated by a single compact component, and there is no
significant variation over the 1.5 years spanned by our
observations. In Fig.~\ref{fig:g351_qso}, we show images of the four
background radio sources at the first epoch.  We fitted
two-dimensional Gaussian brightness distributions to the four
background radio sources and the two brightest maser spots at \VLSR\ =
41.9 \kms\ and 41.5 \kms.  In Fig.~\ref{fig:g351_para}, we plot the
positions of these maser spots relative to the two background radio
sources with superposed curves representing the model maser tracks
across the sky.

Our parallax estimates, given in Table~\ref{tab:g351_para}, for three
of the four background sources are consistent for an individual maser
spot. The parallax estimate using \Jthree\ differs from the others by
about 2-sigma, but gives the smallest post-fit residuals.  Given the
limited number of degrees of freedom for the fits, it is unclear
whether the data for this background source is worse than for the other
three background sources and has a fortuitously good fit, or if it is
the best data and the data for the other three sources agree
fortuitously well.  Rather than make this decision, we use the data
from all four background sources.

The difference in parallax between the two maser spots is roughly 0.10
mas.  The scatter among the 4 measures for each spot suggests an
uncertainty of about 0.03 mas, the joint error in the difference would
be about 0.04 mas.  So the discrepancy is just over 2-sigma. Changes in
maser structure might be a possible cause. The 41.9 and 41.5
\kms~masers varied by 3 and 6 Jy over the 4 epochs. However, with our
limited data we cannot really evaluate which maser gives a better
parallax.  So we have used both and conservatively expanded the
parallax uncertainty to half the difference between the individual
values.

We adopt the ``combined'' parallax solution, which uses both maser
spots and the four background sources, which gives a parallax for
\Gone\ of $0.306 \pm 0.045$~mas.  As done for source \Gzero, we do not
use the formal parallax uncertainty of $\pm0.015$~mas, but instead
conservatively adopt a parallax uncertainty of $\pm0.045$~mas, which
allows for the larger than expected difference in parallax estimates
from the two maser spots.  The parallax for \Gone\ corresponds to a
distance of $3.27^{+0.56}_{-0.42}$~kpc.

The proper motion estimates for the two maser spots and for all four
background sources appear consistent within their joint uncertainties
and the average proper motion is $-0.71 \pm 0.05$~\masy\ toward the
East and $-3.61 \pm 0.17$~\masy\ toward the North.  For the measured
parallax, the proper motion corresponds to $-11.0$~\kms\ and
$-56.0$~\kms\ eastward and northward, respectively. The average LSR
velocity of the spots is 41.7 \kms, which corresponds to a heliocentric
radial velocity of 24.8 \kms.

\begin{table}[H]
  \caption{\Gone\ Parallax \& Proper Motion Fits}
  \label{tab:g351_para}
  \begin{center}
  \begin{tabular}{ccccc}
\hline\hline
Maser \VLSR & Background  & Parallax     &  $\mu_{x}$    &  $\mu_{y}$ \\
(\kms)      & Source      &  (mas)       &  (\masy)      &  (\masy)   \\
\hline
 41.9 ...   & \Jfive   & 0.272 $\pm$ 0.048  &  $-0.76 \pm 0.08$  & $-3.83 \pm 0.66$ \\
 41.9 ...   & \Jthree  & 0.220 $\pm$ 0.021  &  $-0.67 \pm 0.04$  & $-3.71 \pm 0.44$ \\
 41.9 ...   & \Jfour   & 0.287 $\pm$ 0.042  &  $-0.76 \pm 0.07$  & $-3.66 \pm 0.36$ \\
 41.9 ...   & \Jseven  & 0.274 $\pm$ 0.067  &  $-0.75 \pm 0.12$  & $-3.85 \pm 0.58$ \\

 41.5 ...    & \Jfive   & 0.368 $\pm$ 0.047  &  $-0.78 \pm 0.08$  & $-3.62 \pm 0.08$ \\
 41.5 ...   & \Jthree  & 0.319 $\pm$ 0.021  &  $-0.70 \pm 0.03$  & $-3.50 \pm 0.21$ \\
 41.5 ...   & \Jfour   & 0.385 $\pm$ 0.043  &  $-0.78 \pm 0.07$  & $-3.44 \pm 0.31$ \\
 41.5 ...   & \Jseven  & 0.363 $\pm$ 0.066  &  $-0.76 \pm 0.12$  & $-3.63 \pm 0.11$ \\
         &          &                    &                    &                  \\
 41.9 ...   & Combined & 0.306 $\pm$ 0.045  &  $-0.74 \pm 0.07$  & $-3.73 \pm 0.24$ \\
 41.5 ...   &          &                    &  $-0.69 \pm 0.07$  & $-3.50 \pm 0.24$ \\
\hline
  \end{tabular}
  \end{center}
\small{Note. --- see table~\ref{tab:g350_para} note.}
\end{table}

\section{Galactic Locations \& 3-dimensional Motions}

In order to study the 3-dimensional motion of the maser sources in the
Galaxy, we convert their radial and proper motions from the equatorial
heliocentric reference frame in which they are measured into a Galactic
reference frame.  A convenient frame is one rotating with a constant
speed at the position of the maser source: ie, a ``local standard of
rest'' for the location of maser.  The methods used to convert to this
frame will be documented in \citet{Reid08b}. We use the IAU recommended
values of $R_{0} = 8.5$ kpc and $\Theta_{0} = 220$ \kms, and the
Hipparcos solar motion values $U = 10.0 \pm 0.40$, $V = 5.2 \pm 0.60$,
and $W = 7.2 \pm 0.40$ \kms \citep{dehnen98}. For this Galactic model,
\Gzero\ has a peculiar motion of $-13$~\kms\ in the direction of
Galactic rotation, $0$~\kms\ toward the Galactic Center and $-8$~\kms\
toward the North Galactic Pole.  For \Gone, we find peculiar motion
component of $-16$~\kms\ in the direction of Galactic rotation,
$1$~\kms\ toward the Galactic Center and $-9$~\kms\ toward the North
Galactic Pole. Thus, both sources are rotating slower than the Galaxy
spins, that is, slower than for a circular orbit for a flat rotation
curve for the Galaxy. Neither source has a significant peculiar
velocity component toward the Galactic center and both are moving
toward the NGP at about $-8$~\kms.


Research on the structure of the galaxy in Nanjing University is
supported by the National Science Foundation of China ({\bf NSFC})
under grants 1062130, 10673024, 10703010 and 10733030, and NBRPC (973
Program) under grant 2007CB815403. Andreas Brunthaler was supported by
the DFG Priority Programme 1177.



\begin{thebibliography}{}

\bibitem[Braz \& Epchtein(1983)]{Braz83} Braz, M.A., Epchtein, N.,
1983, A\&AS, 54, 167
\bibitem[Brunthaler et al.(2009)]{Brunthaler08} Brunthaler, A. et al. (2009), this volume, (Paper V)
\bibitem[Caswell et al.(1983)]{Caswell83} Caswell, J.L., Haynes,
R.F. 1983, Aust. J. Phys., 36, 417
\bibitem[Caswell et al.(1995)]{caswell95} Caswell, J. L., Vaile,
  R. A., Ellingsen, S. P., Whiteoak, J. B. \& Norris, R. P., 1995, \mnras,
  272, 96
\bibitem[Churwell et al.(1990)]{Churchwell90}
  Churchwell, E., Walmsley, C.M., Cesaroni, R., 1990, 159, A\&AS, 83, 119
\bibitem[De Buizer et al.(2004)]{buizer04} De Buizer, J. M.,
  Radomski, J. T., Telesco, C. M. \& Pinal, R. K., 2004, \aaps, Vol. 36,
  1619
\bibitem[De Buizer et al.(2006)]{buizer06} De Buizer, J. M, James M. , 2006
  \apj, 642L, 57
\bibitem[Dehnen \& Binney(1998)]{dehnen98} Dehnen, W., \& Binney, J. J.,
  1998, \mnras, 298, 387
\bibitem[Fomalont et al.(2003)]{Fomalont03} Fomalont, E., Petrov, L., McMillan, D.
  S., Gordon, D., \& Ma. C., 2003, \aj, 126, 2562
\bibitem[Forster \& Caswell(1989)]{forster89} Forster, J. R., Caswell,
  J. L., 1989, \aap, 213, 339
\bibitem[Genzel \& Downes(1977)]{Genzel77} Genzel, R., Downes, D.,
1977, A$\&$AS, 30, 145
\bibitem[Gibb et al.(2003)]{gibb03} Gibb, A. G., Hoare, M. G., Little,
  L. T., Wright, M. C. H., 2003 \mnras, 339, 1011
\bibitem[Hachisuka \etal (2006)]{Hachi06}
  Hachisuka, K. \etal\ 2006, \apj, 645, 337
\bibitem[Kalenskii et al.(2000)]{kalenskii00} Kalenskii, S. V.,
  Promislov, V. G., Alakoz, A., Winnberg, A. V. \& Johansson, L. E. B.,
  2000 \aap, 354, 1036
\bibitem[Palagi et al.(1993)]{Palagi93} Palagi, F., Cesaroni, R.,
Comoretto, G., Felli, M., Natale, V., 1993, A\&AS, 101, 153
\bibitem[Petrov et al.(2005)]{Petrov05}Petrov, L., Kovalev, Yu. Y., Fomalont,
E., \& Gordon, D., 2005, \aj, 129, 1163
\bibitem[Ma et al.(1998)]{Ma98} Ma C., Arias E. F., Eubanks T. M., et al., 1998, \aj, 116, 516
\bibitem[Moscadelli et al.(2009)]{Moscadelli08} Moscadelli, L. et al. 2008, this volume, (Paper II)
\bibitem[Reid et al.(2009)]{Reid08a} Reid, M. J. et al. (2009) this volume, (Paper I)
\bibitem[Reid et al.(in preparation, Paper VI)]{Reid08b} Reid, M. J. (in preparation), Paper VI
\bibitem[Solomon et al.(1987)]{solomon87} Solomon, P. M., Rivolo,
A. R., Barrett, J., Yahil, A., 1987 \apj, 319, 730
\bibitem[Vallee \& MacLeod(1990)]{Vallee90} Vallee, J.P., MacLeod,
J.M., 1990, ApJ, 358, 183
\bibitem[Xu et al.(2006a)]{Xu06a} Xu Y., Reid M. J., Zheng X. W. \&
  Menten, K. M. 2006a, {\it Science}, 311, 54
\bibitem[Xu et al.(2006b)]{Xu06b} Xu Y., Reid M. J., Menten K. M., Zheng X. W., 2006, \apjs, 166, 526
\bibitem[Xu et al.(2009)]{Xu08} Xu, Y. et al. (2009), this volume, (Paper III)
\bibitem[Zhang et al.(2009)]{Zhang08} Zhang, B. et al. (2009), this volume, (Paper IV, this paper)

\end{thebibliography}
\end{document}